# All-BN Distributed Bragg Reflectors Fabricated in a Single MOCVD Process

*Arkadiusz Ciesielski[1], Jakub Iwański[1], Piotr Wróbel[1], Rafał Bożek[1], Sławomir Kret[2], Jakub Turczyński[2], Johannes Binder[1], Krzysztof P. Korona[1], Roman Stępniewski[1], Andrzej Wysmołek[1]*

1. University of Warsaw, Faculty of Physics, Pasteura 5, 02-093 Warsaw, Poland
2. Institute of Physics, Polish Academy of Sciences, Al. Lotników 32/46, 02-668 Warsaw, Poland

**Abstract**

Distributed Bragg Reflectors (DBR) are well-established photonic structures that are used in many photonic applications. However, most of the DBRs are based on different materials or require post-process etching which can hinder integration with other components in the final photonic structure. Here, we demonstrate the fabrication of DBR structures consisting only of undoped boron nitride (BN) layers with high refractive index contrast by using Metal-Organic Chemical Vapor Deposition (MOCVD). This has been achieved in a single process, without the need for any post-process etching. The difference in the refractive index of the component BN layers stems from different degrees of porosity of the individual BN layers, which is a direct result of a different growth temperature. The fabricated DBR structures consist of 15.5 pairs of BN layers and exhibit a reflectance of 87±1% at the maximum. The wavelength of maximum reflectance can be tuned from 500 nm up to the Infrared Region (IR), by simply adjusting the growth periods of subsequent BN layers. We also demonstrate that the fabricated structures can be used to create an optical microcavity. The fabricated DBRs are very promising candidates for future applications, for example in combination with single-photon emitters in h-BN, which could allow the building of a cavity-based all-BN single-photon source.

## 1. Introduction

Multiple thin films with significant refractive index contrast may be stacked on top of each other to create a Distributed Bragg Reflector (DBR) – a one-dimensional photonic crystal with a photonic bandgap in the direction perpendicular to the layer planes. Such structures exhibit reflectance values approaching 100% for photon energies within the bandgap. Depending on the refractive index values of the incorporated thin films, their thicknesses and the number of layer pairs, the position and width of this maximum can be controlled [1]. DBRs find extensive use primarily in vertical-cavity surface-emitting lasers (VCSEL) [2-6], resonant-cavity light-emitting diodes (RCLED) [7,8], but also in photovoltaics [9,11] and diffractive optical components [12]. MOCVD is already an established growth method for DBR fabrication, as the first reports date back to the mid-80s [13, 14]. However, DBR structures based on boron compounds have been explored only recently [15,16].

Boron nitride (BN) crystals, and in particular those which exhibit $sp^2$ electronic orbital hybridization – hexagonal and rhombohedral BN are promising candidates to use in thin-film applications, in flexible electronics [17], as substrates for other 2D materials [18, 19], or as single-photon sources [20-22]. These BN polymorphs exhibit a wide optical bandgap of approximately 6 eV [23, 24]. High-quality $sp^2$-type BN layers are commonly grown using the MOCVD technique with trimetylborane (TMB) or triethylborane (TEB) as the boron source and ammonia as the nitrogen source [25-28]. Recently, flow modulation epitaxy of such films is becoming more common [29]. To investigate many of their properties, it is necessary to insert them into the microcavity of the DBR structures in order to increase the number of photons interacting with the layer. This requires delamination of the $sp^2$-type BN layers from the original substrate onto the DBR. Although there have been successful attempts at delamination of hBN films [30], the process is still far from perfect and induces changes to fabricated material.

The most widely used DBRs are based on different materials that provide a large refractive index contrast. However, such an approach can have drawbacks in terms of integration with other materials in the final photonic structure, for instance due to the inherent difference in thermal expansion of the different materials, or the requirement of lattice matching for epitaxial structures. Another approach to obtain the needed reflective index contrast is to use porous structures. However, such structures require a selective post-fabrication etching process [31-34].

In this work, we present a different approach. We implemented a low-temperature, low-pressure growth regime [35] to fabricate DBR structures entirely based on BN layers with high refractive index contrast. We achieve this in a single process by only changing the growth temperature of the BN layers. We also show that the fabricated DBRs may easily be used to create a microcavity. This gives hope for future growth of all-BN single-photon sources, where the active area made of hBN is grown directly on top of a BN-based DBR and covered with another BN-based DBR in a single MOCVD process.

## 2. Materials and methods

Boron nitride layers were obtained by MOCVD using an Aixtron CCS 3 × 2" reactor. Growth was performed on single-side polished, 430 μm-thick, two-inch sapphire c-plane wafers (with up to 0.2º misorientation) as substrates, except for the sample discussed in section 3.3, where a 330μm-thick double-side polished substrate was used. Ammonia and triethylborane (TEB) were used as precursors for nitrogen and boron, respectively. The carrier gas was set to hydrogen for most samples, but for selected samples nitrogen was used. We note that in the text where it is the case. The system temperature was controlled via the ARGUS multipoint pyrometer. Since these two measurement techniques produce slightly different results, the values from the ARGUS pyrometer will be reported as the growth temperature. The system is also equipped with a monochromatic optical reflectometer, capable of measuring normal reflectance from the sample's surface at the wavelength of 635 nm. This allows to estimate the thickness and refractive index (at 635 nm) of the fabricated BN layers.

All of the BN layers have been fabricated at a chamber pressure of 100 mbar, TEB flow of 80 ccm liters per minute and ammonia flow of 100 ccm. The only free parameter was the system temperature. In the case of the DBR samples, where two types of BN layers were grown at different temperatures, the flow of precursor gases was interrupted as the system was ramping or cooling to the desired temperature.

Ellipsometric measurements were carried out with the use of the Woollam RC2 dual rotating compensator ellipsometer with a vertical auto angle stage. This architecture allows for the measurement of the full Mueller Matrix as well as the depolarization factor. The samples were prepared on roughened sapphire substrates that ensure a lack of depolarization due to reflection from the bottom side of the substrate and simplify ellipsometric analysis since in this case, the measurement is sensitive only to the in-plane component of the substrate's refractive index. Samples were measured in reflection mode in the wavelength range from 190 nm to 1700 nm for several angles of incidence in the range from 40º to 75º to improve precision. Modeling and fittings were done with the use of CompleteEASE v6.61 software. The optical model took into account the semi-infinite sapphire substrate of refractive index measured prior to the main experiment, and the BN layer was parametrized with a generalized oscillator approach with the use of one or more Tauc-Lorentz and Lorentz oscillator models [36]. To retrieve optical constants of the BN layers multisample analysis was carried out where the fitting of the model was done to several samples simultaneously while geometrical parameters like thickness or roughness were extracted for subsequent samples.

The normal reflectance spectra were recorded using Ocean Optics HR4000CG UV-NIR Spectrometer with a mercury lamp as a light source. Presented spectra are an average result over 25 acquisitions.

AFM measurements were performed in the tapping mode using a Multimode AFM (MMAFM-2) with a Nanoscope IIIa controller from Bruker (formerly Digital Instruments). The probes used were NT-MDT NSG30 with a guaranteed tip radius of 10 nm.

Specimens for cross-sectional transmission electron microscopy (TEM) measurements were prepared using a standard method of mechanical pre-thinning followed by Ar ion milling at 4KV and 500V acceleration voltage of Ar ions. TEM characterization was performed using an image corrected Titan Cubed 80–300 microscope operating at 300 kV. The scanning transmission electron microscopy was performed using an annular dark-field (HAADF) Fischione 3000 detector. The elemental composition determination was performed by energy-dispersive x-ray spectroscopy (EDX) using EDAX 30 mm2 Si(Li) detector with a collection angle of 0.13 srad.

Numerical modeling of the optical parameters of the fabricated structures was performed using the Transfer Matrix Method (TMM) implemented as MATLAB code, similarly as in other works [37]. In the calculations, the refractive index values for sapphire were taken from the literature [38]. Effective Medium Approximation calculations have been performed using the Bruggeman approach [39]. A detailed description can be found in the supplementary material.

## 3. Results and Discussion

### 3.1 Growing BN layers with high refractive index contrast

We have fabricated two kinds of BN layers to serve as component layers for the Distributed Bragg Reflector. One of them has been grown at 640 ºC system temperature, the other at 820 ºC. We kept all other growth parameters constant. Figure 1 shows the in-situ evolution of the sample reflectance at 635 nm, during the deposition of BN layers using hydrogen as the carrier gas, grown at 640 ºC (Fig 1a) and 820 ºC (Fig 1b). For the layer grown at the lower temperature, the reflectance increases at first, reaching a maximum of 11.5%, then drops to the level of pure sapphire. For the layer grown at the higher temperature, it is the reverse – at first, the reflectance decreases, reaches 5.9%, then increases to the level of pure sapphire. This means that both of the fabricated BN layers are transparent at a wavelength of 635 nm. However, since we observed opposite extrema for these two samples, it means that one of them has a higher refractive index than sapphire, and the other one has lower. A simple TMM model of a single BN layer on a sapphire substrate allows to plot the reflectance as a function of layer thickness for a given refractive index value. This is represented by blue curves in Fig. 1. Best fits to the in-situ measurement were achieved assuming the refractive indices of the grown BN layers are 1.890 and 1.702 for the samples grown at 640 ºC and 820 ºC, respectively. The small discrepancies between the measured and simulated curves result most likely from the slightly non-uniform growth rate of the BN layers. Apart from the refractive index values, the same calculation allows to determine the thickness of the fabricated BN layers. The full Fabry-Perot cycles for refractive indices of 1.890 and 1.702 should occur at 168 nm and 186.5 nm, respectively. From our simulations, we get thickness $d = 167$ nm for the sample grown at 640 ºC and $d = 181$ nm for the sample grown at 820 ºC. This means, that for both layers, a nearly full Fabry-Perot cycle has occurred.

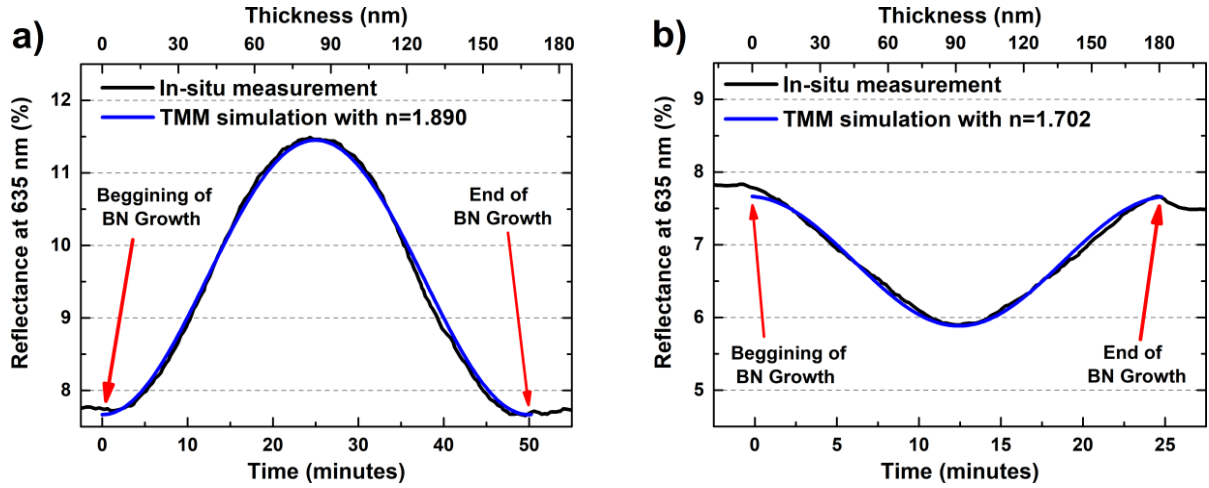

Fig. 1. In-situ reflectance evolution (black curves) of sample BN layers deposited using hydrogen as the carrier gas, at a) 640 °C and b) 820 °C, as a function of the growth time. The red arrows mark the beginning and end of the BN growth process. Blue curves represent simulated reflectance values of a layer with a given refractive index deposited on a sapphire substrate, as a function of that layer thickness.

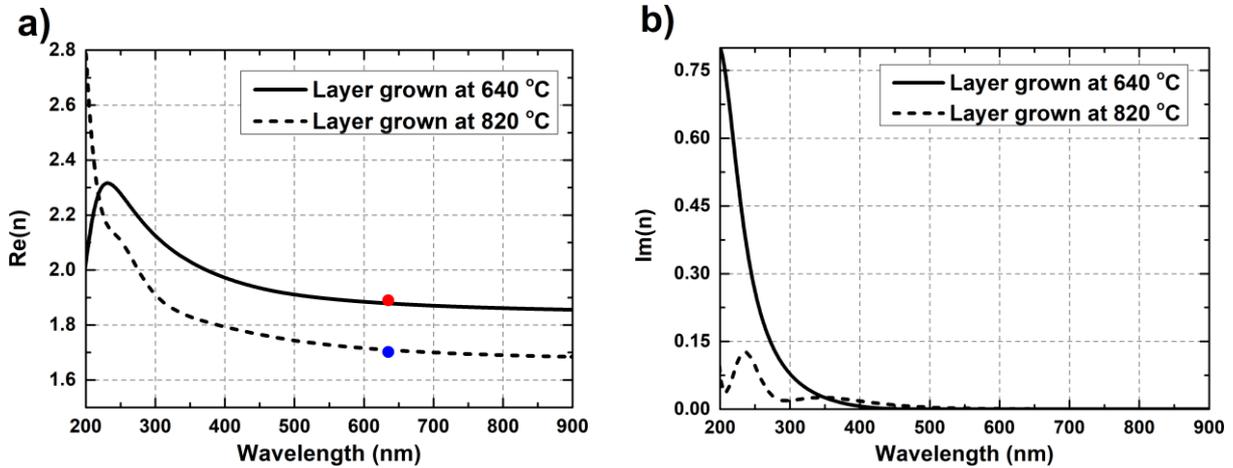

Fig. 2. a) real and b) imaginary parts of the refractive index of the BN layer grown at 640 °C (continuous curves) and 820 °C (dashed curves) determined using spectroscopic ellipsometry. The red and blue dots in a) represent results from in-situ reflectometry for BN layers grown at 640 °C and 820 °C, respectively.

More information about the optical properties of the samples can be acquired using spectroscopic ellipsometry. Optical constants were extracted from ellipsometric azimuths Ψ and Δ measured in the experiment with the use of the model described in the Materials and Methods section. Since the measurement is performed in a broad range of wavelengths, we can determine the refractive index dispersion of the investigated layers. This is presented in Figure 2. At 635 nm, the refractive index values of the fabricated layers determined using spectroscopic ellipsometry are in good agreement with the values calculated from the in-situ reflectance profile – discrepancies do not exceed 0.7%. For both BN layers, the refractive index for wavelengths above 500 nm may be treated as slowly varying, therefore the refractive index contrast is almost constant, at an average of 0.17. Furthermore, for this wavelength range, the imaginary part of the refractive index is negligible, therefore the layers may be treated as non-absorbing. As such, these layers are suitable for the construction of DBR structures optimized for wavelengths greater than 500 nm. Below that value, the imaginary part of the refractive index becomes significant. In the case of the BN grown at 640 °C, the imaginary part of the refractive index rises rapidly with decreasing wavelength, reaching 0.8 at 200 nm. The main absorption resonance in the permittivity spectrum occurs at wavelengths around 200 nm in the case of the BN layer grown at 640 °C, which is in good agreement with results in [40] for hBN and tBN. In the case of the layer grown at 820 ºC, the main absorption resonance occurs for wavelengths below 200 nm, which is indicated by the monotonic behavior of the real part of the permittivity within the whole 200 nm – 900 nm wavelength range. Moreover, an additional resonance at 5.28 eV occurs, which was not observed in [40].

The real part of the refractive index of bulk BN is close to 2.3 for visible wavelengths [41]. In the case of both of our BN layers, this value is much smaller. This suggests that both types of layers are at least partially amorphous or perhaps porous, and to a different extent. Figure 3 presents STEM micrographs of both investigated BN layers. The scans show areas of distinctively higher (brighter parts of the graph) and lower (darker parts of the graph) density in both types of BN layers. This indicates that both layers are indeed porous. The lower density areas in the case of the BN layer grown at 640 ºC (Fig 3a) are smaller and more finely distributed than in the case of the BN layer grown at 820 ºC (Fig 3b). This implies that the latter exhibits higher porosity than the former. Since the pores are filled with either vacuum or air (with refractive index close to 1), the layer with higher porosity should exhibit a lower refractive index, which is exactly what we observe.

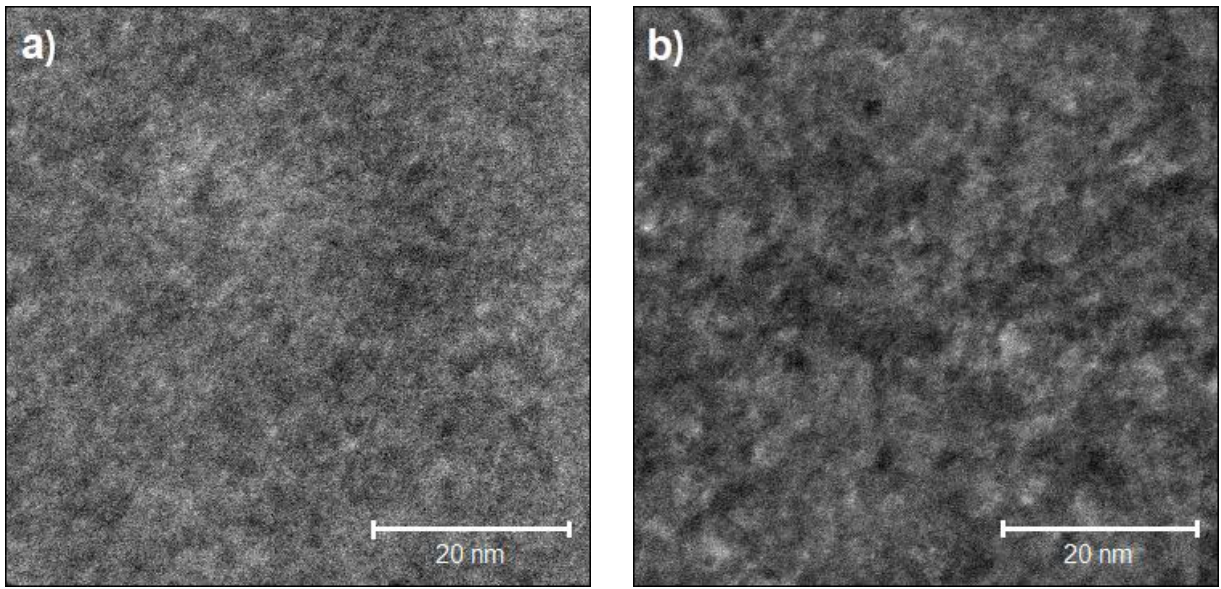

Fig. 3. Cross-sectional STEM micrographs of BN layers grown at a) 640 ºC and b) 820 ºC. Brighter parts of the graph represent areas of larger density, while darker parts represent areas of lower density.

### 3.2 Fabrication and characterization of Distributed Bragg Reflectors

DBRs with the best performance (i.e. highest and broadest reflectance maximum for a given number of layer pairs) are achieved when the component layers have a thickness $d = \lambda/4n$ where $\lambda$ is the wavelength for which the reflectance maximum has to occur and $n$ is the real part of refractive index [1]. For DBRs component layers with refractive indices of 1.89 and 1.702, and the wavelength of 635 nm, the optimal layer thickness should be 84.0 and 93.3 nm, respectively. However, since the refractive index contrast between the constituent BN layers is so high, small changes in individual layer thickness should result only in the shift of wavelength for which the maximum reflectance is observed. The value of maximum reflectance should not change drastically with the thickness of each type of layer, as long as the position of the maximum is at a wavelength of 500 nm or greater. Detailed simulations regarding this matter can be found in the supplementary material.

We fabricated four DBR structures using two approaches. Samples A and B have been fabricated with constant growth times of individual BN layers to ensure similar thicknesses among BN layers grown at the same temperature. In the case of samples C and D, growth times have been manually adjusted during the MOCVD process, so that switching between the growth of high and low refractive index BN layers occurs exactly at the reflectance extremum for 635 nm, thus ensuring that the maximum reflectance will be reached at this particular wavelength. Variable process parameters for all DBRs are specified in Table 1. For each DBR, we have chosen to fabricate 15.5 layer pairs to achieve a reflectance peak in the vicinity of 90% with a clearly defined maximum (see Fig. F1 in the supplementary material for more details). Schematically, such a structure is presented in Fig 4a. Figure 4b presents an AFM amplitude image from a cross-section of the DBR A. A clear layered structure of the sample is observed. This proves that we have indeed fabricated DBR-like layered structures. The pseudocolor contrast between the two types of BN layers on the AFM scan results from the fact, that in the case of the more porous BN layers, the AFM tip would make a deeper recess than in the case of a less porous layer, before the same force acts upon it. It is also worth noting, that for improved visibility of the layer borders, the AFM scan was conducted at a slight angle with respect to the cross-section.

| DBR sample | Time of BN growth at 640 ºC [min.] | Time of BN growth at 820 ºC [min.] | Number of BN layer pairs | Carrier gas |
|---|---|---|---|---|
| A | 27.5 | 14.5 | 15.5 | Hydrogen |
| B | 26.5 | 16.25 | 15.5 | Hydrogen |
| C | 24.2±0.9 | 15±0.4 | 15.5 | Hydrogen |
| D | 6±0.2 | 5.5±0.2 | 15.5 | Nitrogen |

Table. 1. Variable process parameters of the fabricated DBR structures. Values for samples C and D are average and may vary depending on the individual BN layer. Growth times of all individual component BN layers of samples C and D are available in table T1 in the supplementary material.

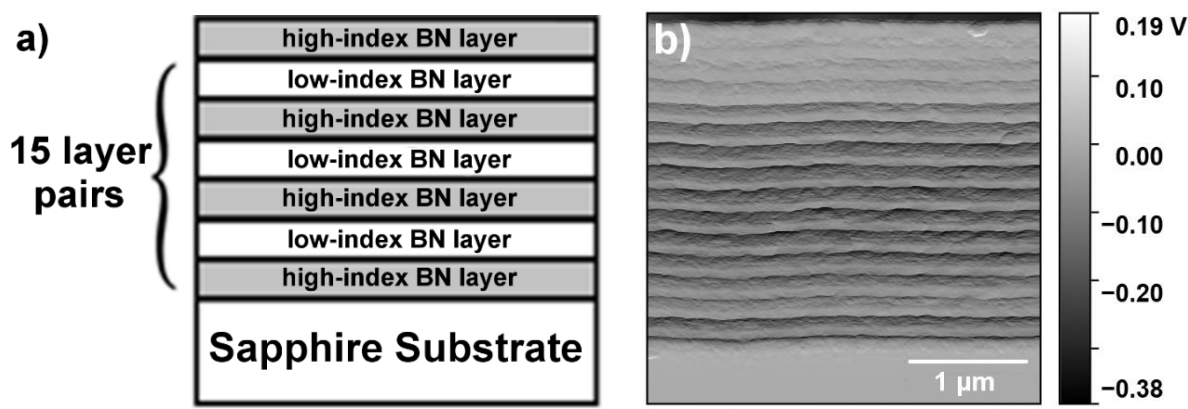

Fig. 4. a) schematic of the DBR; b) AFM amplitude image of a cross-section of DBR A, collected in tapping mode.

Figure 5a presents the in-situ reflectance evolution of the growing DBRs A and C. In the case of DBR A, for which the growth times of each individual BN layer have been fixed, not much information can be extracted from this graph – during the growth of the DBR, it is unclear what is the wavelength for which the reflectance maximum will occur. In the case of DBR C, manual control of the growth times allowed for switching between growing high and low refractive index BN layers exactly at the reflectance extrema. This ensured that the reflectance maximum will occur close to 635 and allowed for monitoring the maximum reflectance value after growing each individual layer pair. Such an in-situ optimized growth process is in principle possible for different wavelengths provided the right equipment is available.

Figure 5b presents the post-process reflectance spectra of the fabricated DBR samples recorded at the center of the 2" wafer on which the DBR was deposited, while Figure 6 depicts the mapped peak reflectance level for the whole 2" DBR wafer in the case of all fabricated DBRs.

15.5 layer pairs were enough to achieve 87±1% reflectance at the maximum, measured at the center of the wafer. For some areas of the wafer that value was greater, exceeding even 90% in the case of DBR C. The spectral width of the high reflectance area for which the reflectance is greater than 80% is about 30 nm.

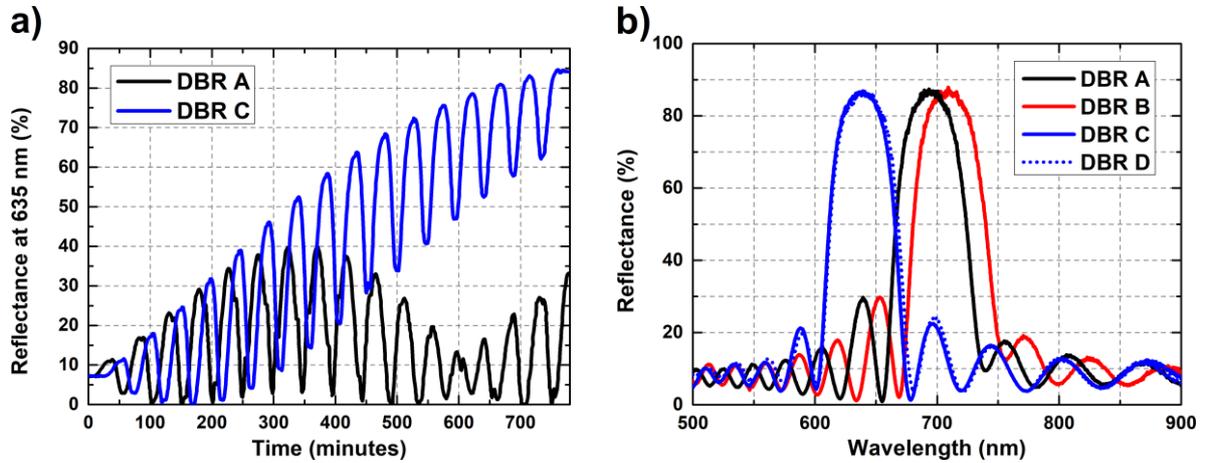

Fig. 5. a) in situ-reflectance spectra of selected DBR structures, measured for 635 nm at the center of the wafer. b) post-process reflectance spectra of the fabricated DBR structures, measured at the center of the wafer.

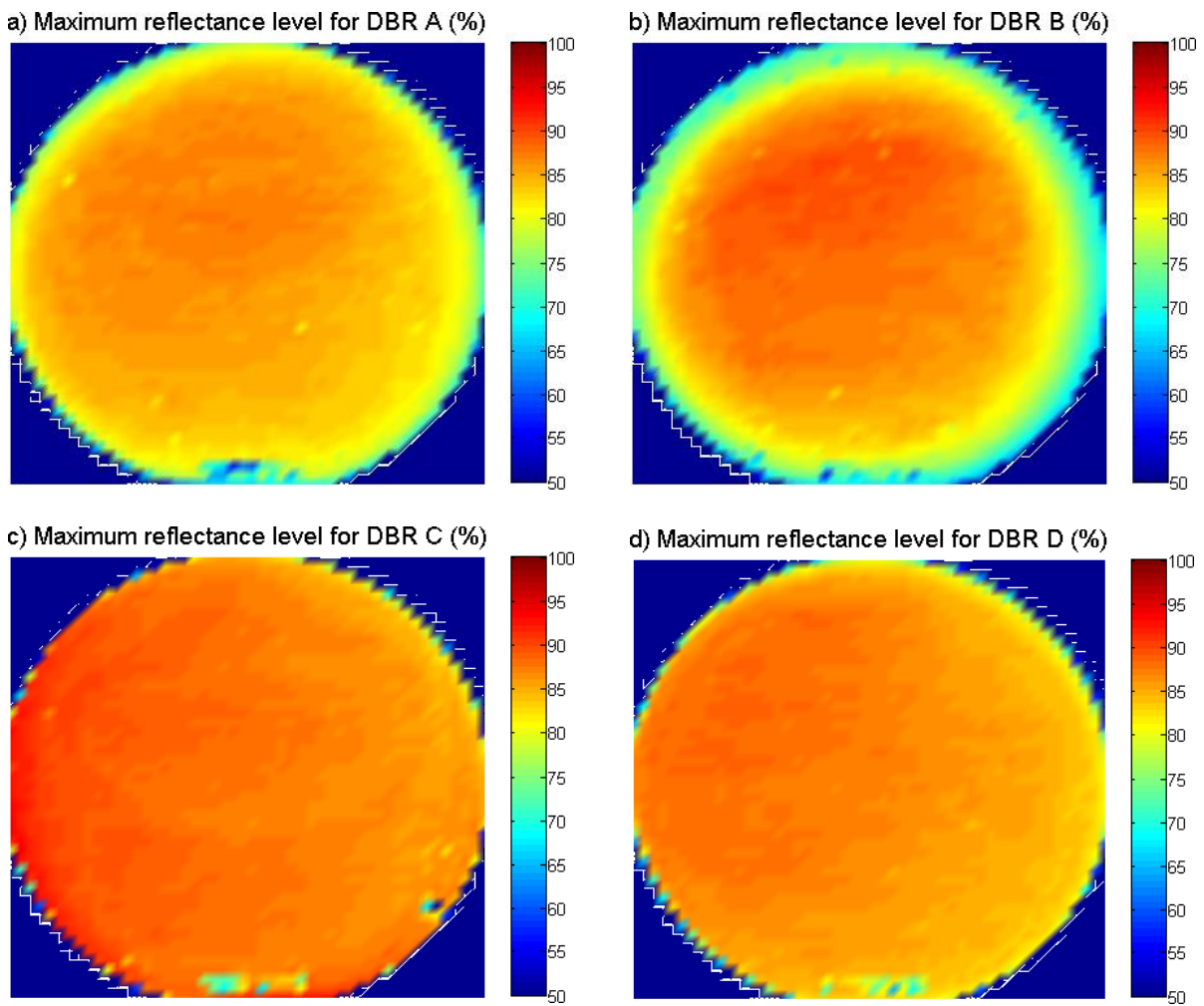

Fig. 6. Peak reflectance level measured at different points of the 2" DBR samples. The upper parts of the wafers were located closer to the center of the showerhead in the MOCVD process. The anomalies in the lower part of the graphs are due to the presence of serial numbers on the substrate wafers.

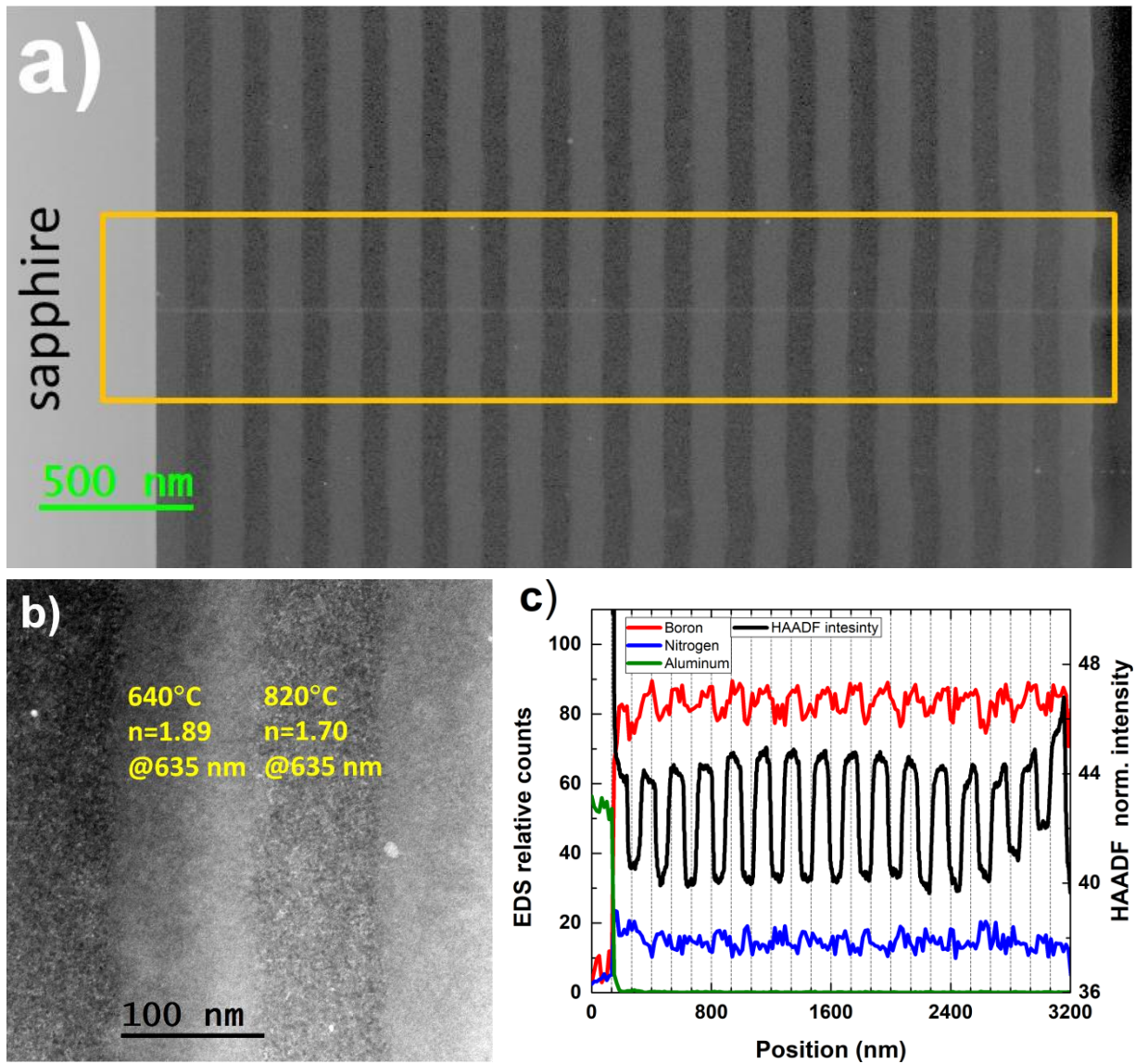

Fig. 7. a) and b) cross-sectional STEM micrographs of DBR A. Brighter layers (more dense) correspond to the high-index BN and darker layers correspond to the low-index material with high porosity. The brightest part on the left-hand side of scan a) corresponds to the sapphire substrate. c) STEM intensity profiles averaged over the orange frame width in graph a), along with EDX signal from boron, nitrogen and aluminum atoms (relative, normalized signal of integrated intensity of peeks in EDX spectra). The bright line inside the frame in a) shows the real position and size of the electron beam during EDS measurement. The variation of the image intensity across individual layers in b) is related to the inhomogeneous thickness of the cross-section due to different rates of milling of dense and porous material or vicinity of glue used in the preparation of the sample for measurement.

Figure 7ab presents STEM micrographs of a cross-section of DBR A. The sample clearly exhibits a layered structure with alternating high-index BN layers (bright parts of the scan) and low-index BN layers (dark parts of the scan). However, the individual layers exhibit waviness and slight variation in their thickness depending on the spot in the sample. These fluctuations are in the sub-micrometer scale and should not affect DBR properties, as is confirmed by further simulations (see Fig. 8). The porosity contrast between neighboring BN layers grown at different temperatures is clearly noticeable. More porous, less dense layers appear darker in the STEM image. The black curve in Figure 7c presents the intensity of the STEM signal extracted across horizontal lines from Fig. 7a, averaged over a 500 nm wide area. The high-intensity areas correspond to the high-index BN layers and the low-intensity areas correspond to the low-Index BN layer, therefore, this figure allows to estimate the thickness of each individual BN layer.

The red and blue curves in Fig. 7c represent the relative EDS signal from boron and nitrogen atoms, respectively. Higher levels of the boron signal coincide with the high level of the STEM signal, while high levels of the nitrogen signal coincide with low levels of the STEM signal. Therefore, we can conclude, that the stoichiometry of the BN is different in the layers grown at different temperatures, which may also contribute to the observed differences in the refractive index values.

We have implemented the thickness values determined from Figure 7 into the TMM model to simulate the reflectance profiles of DBRs A and B. The refractive index values of both types of BN layers were taken from Fig 2. The exact simulation procedure is described in the supplementary material. The results for DBR A are presented in Figure 8a. Although there are some discrepancies between the measured and simulated curves, the fit reproduces the shape of the reflectance curve very well. Similar calculations were performed in the case of DBR B (Fig 8b). In this case, the exact thicknesses of the BN layers were not known. However, a very good fit was achieved by assuming, that the ratio of individual layer thickness between DBRs A and B is equal to the ratio of the growth times for these layers.

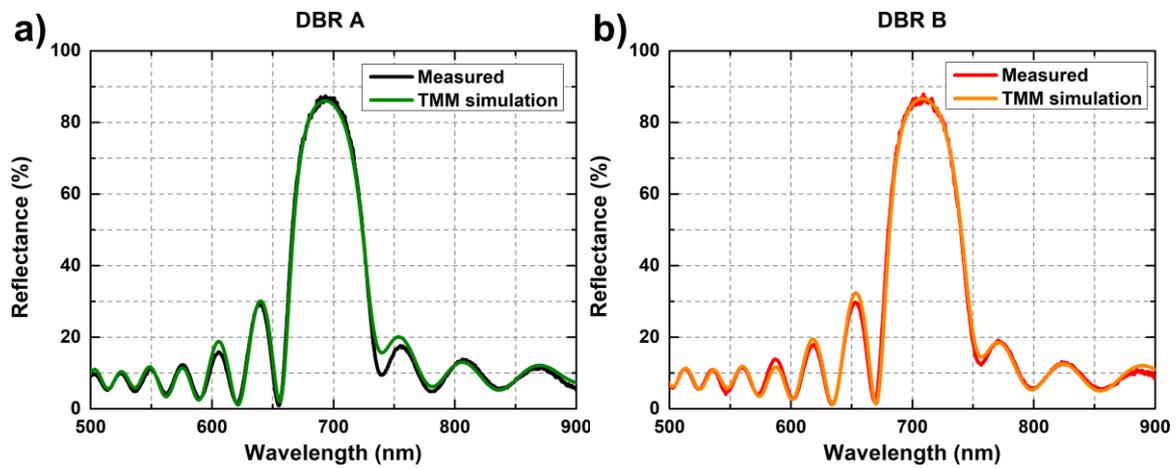

Fig. 8. TMM fitted spectral reflectance curves for a) DBR A and b) DBR B.

## 3.3 Creating microcavities from the BN-based DBRs

To construct a microcavity, we have fabricated yet another DBR structure (DBR E), however this time, it was deposited onto a double-side polished sapphire substrate instead of a single-side polished one. The growth times of both types of BN layers have been adjusted so that the reflectance maximum at the center of the wafer would occur at around 650-660 nm. This spectral region is particularly interesting in case of 2D materials such as $MoS_2$ since its exciton energy corresponds to this spectral range. The growth details for DBR E are listed in Table 2. Reflectance maps for this DBR are available in the supplementary material.

| Time of growth for first high-index BN layer [min.] | Time of growth for next high-index BN layers [min.] | Time of growth for low-index BN layers [min.] | Number of BN layer pairs | Carrier gas |
|---|---|---|---|---|
| 6.916 | 6.25 | 5.716 | 15.5 | Nitrogen |

Table. 2. Process parameters of the new DBR E.

An optical microcavity is a space inserted between two DBR structures. To achieve the construction of such a cavity, we have deposited a non-transparent layer of aluminum at the edge of the wafer with the DBR E, then cut the wafer in half. The resulting structure is presented in Figure 9a. We have then placed one half of the wafer on top of the other in such a way that the aluminum layers are in direct contact. Lastly, we have placed additional weights to press down the edges on opposite sides of the wafer (Fig. 9b). That way, the aluminum serves as a spacer and allows to create a wedge-like air-filled microcavity between two identical DBRs. Figure 9c depicts a schematic drawing of the mapped area, with selected pixels marked with different colors. Figure 9d presents the reflectance profiles measured at the marked points. Movie V1 in the supplementary material presents the reflectance measured for all points on this line. Constructing even such a simple system allows observing the wedge-like microcavity. Near the aluminum spacer, the cavity is too large to observe any dip in the reflectance at the high-reflectance plateau. As the distance from the spacer increases, the width of the air cavity decreases. In turn, two dips in the reflectance spectrum arise, then transition into one dip, then back to two. The position of these dips also varies depending on the spot, which is also a result of the different width of the air cavity. The Q-factor of the cavity measured at the spot marked in black in Fig 9c equals 220. Using TMM calculations, we simulated the spectral reflectance curves of the DBR optimized for the wavelength of 650 nm with an air cavity of varying thickness (see Video V2 in the supplementary material). Based on these simulations, we estimate the thickness of the acquired cavity to be in the range of 3-7 microns.

Due to the easy construction, many 2D materials such as transition metal dichalcogenides can be inserted in such microcavities. In theory, this provides means to measure weak optical signals from these materials or even single-photon phenomena. Another interesting approach would be to fabricate an all-BN single-photon source, by depositing an appropriate hBN layer [20,21], possibly with intentional carbon defects [22] on top of a BN-based DBR followed by depositing another BN-based DBR in a single MOCVD process.

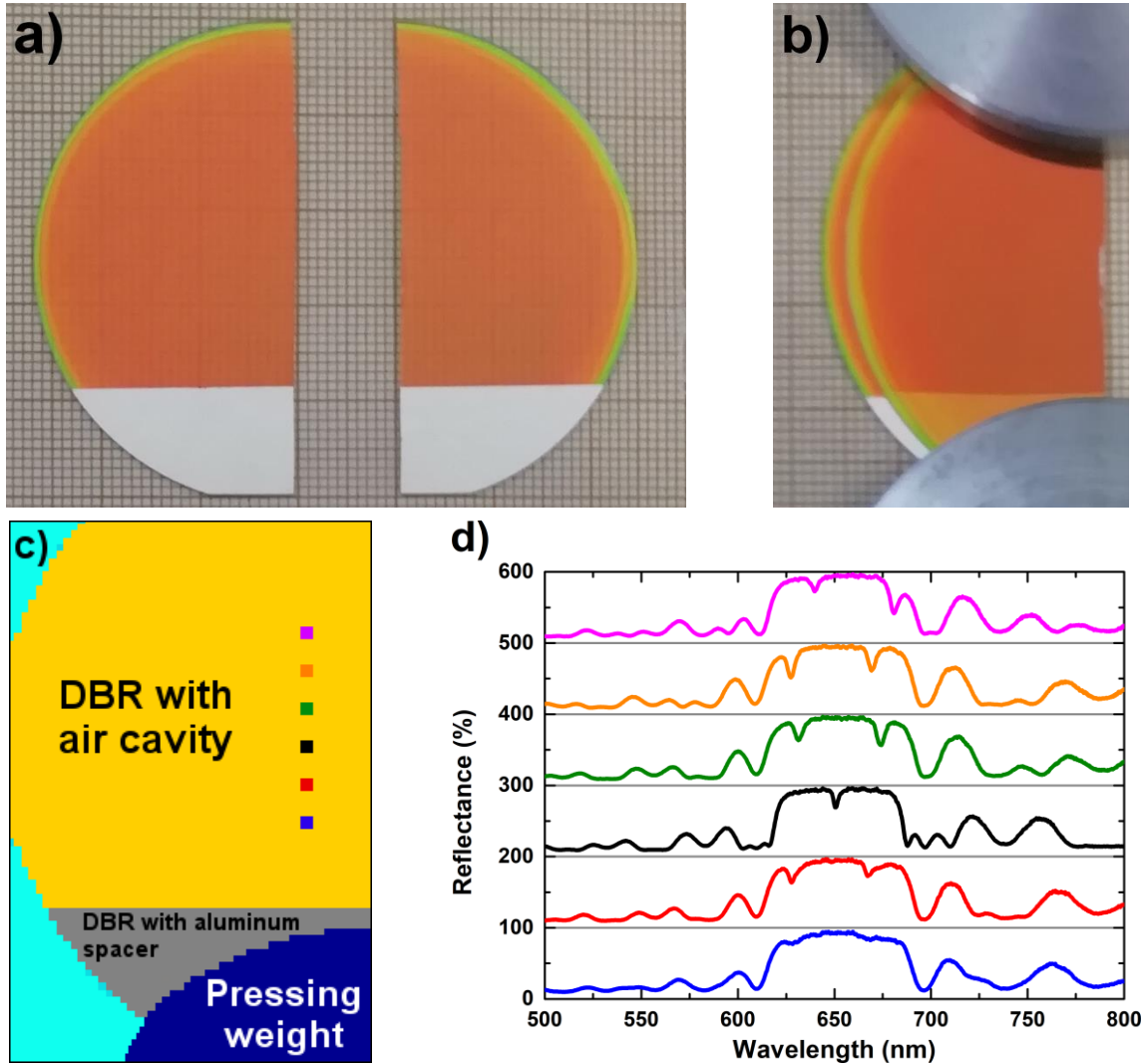

Fig. 9. a) photograph of DBR E with aluminum spacer deposited near the edge of the wafer, b) DBR microcavity set up as for measurement, c) schematic of the mapped area, d) reflectance profiles measured at the points marked in c). The colors of the curves correspond to the colors of the points. Each curve has been offset by 100% with respect to the previous one.

## 4. Conclusions

Using the MOCVD technique, we have fabricated DBR structures consisting purely of BN. To that end, we have fabricated two types of BN layers with high refractive index contrast by modifying only the system temperature during the growth process. STEM results have shown that both types of fabricated layers are porous but to a different extent. This difference in porosity is most likely behind the difference in the refractive indices of these layers. Implementing thickness data extracted from the STEM results as well as refractive index values from spectroscopic ellipsometry into a Transfer Matrix Model, we were able to achieve a good numerical fit to the experimental normal reflectance data. We have also shown how to utilize the fabricated DBRs to construct an optical microcavity.

**Acknowledgments**
This work was funded by the National Science Centre, grants no. 2019/33/B/ST5/02766 and 2020/39/D/ST7/02811.

# All-BN Distributed Bragg Reflectors Fabricated in a Single MOCVD Process

## SUPPLEMENTARY MATERIAL

Figure F1a presents TMM simulated reflectance profiles of DBR structures consisting of the two types of fabricated BN layers (n=1.89 and 1.702 at the wavelength of 635 nm) with thicknesses optimized so that the reflectance maximum occurs at 635 nm, for different numbers of layer pairs. 15.5 layer pairs are enough to achieve peak reflectance reaching almost 90% with a clearly defined maximum. Fabricating a DBR with more layer pairs is beneficial from the engineering point of view, but provides little more insight into the physics of the structures. Figure F1b presents TMM simulated reflectance profiles of DBR structures with 15.5 layer pairs each, but with layer thickness optimized so that the reflectance maximum occurs at different wavelengths. For wavelengths greater than 500 nm, the peak reflectance remains similar, while for shorter waves, it decreases significantly with the wavelength, due to the increasing value of the extinction coefficient. Therefore, BN layers deposited using our procedure are not suitable to fabricate DBRs designed for UV, as opposed to [1,2].

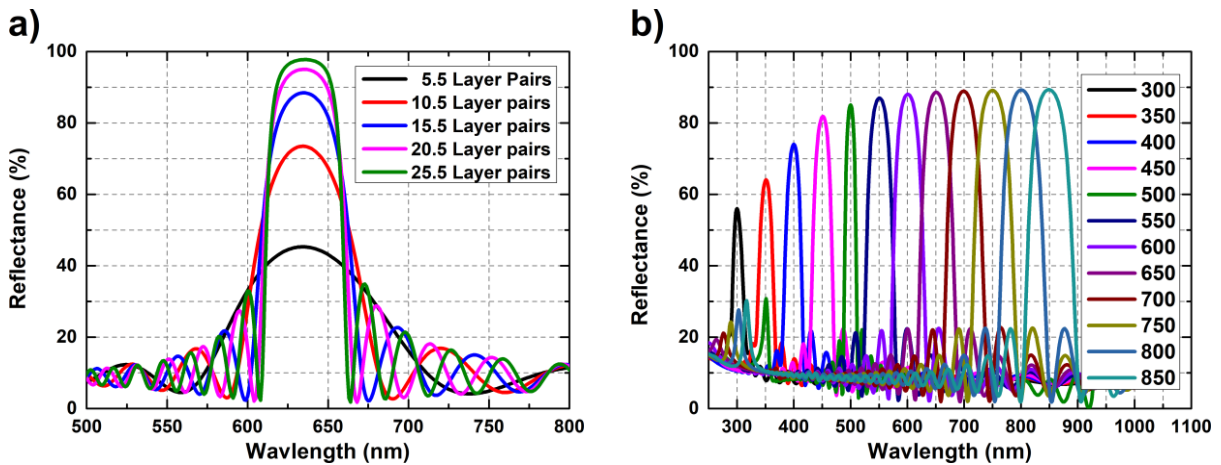

Fig. F1. Simulated reflectance profiles of DBR structures a) optimized for 635 nm and consisting of a different number of layer pairs; b) consisting of 15.5 layer pairs and optimized for different wavelengths in the 300-800 nm range.

Table T1 presents the exact growth times of each individual BN layer, for DBRs C and D. As the growth times were manually adjusted so that the switch between the growth of high- and low-index BN layers occurs exactly at the reflectance extremum, these values differ slightly between each layer. It is worth noting, however, that for both the growth times of the very first BN layer are significantly larger than for all other layers. This indicates, that for the first BN layer, the interaction between the growing BN and the substrate inhibits the layer growth. For subsequent layers, this interaction is much smaller, due to larger separation from the substrate.

| DBR C | | DBR D | |
|---|---|---|---|
| Growth time of subsequent high-index BN layers (min.) | Growth time of subsequent low-index BN layers (min.) | Growth time of subsequent high-index BN layers (min.) | Growth time of subsequent low-index BN layers (min.) |
| 26.67 | 15.00 | 6.67 | 5.33 |
| 25.50 | 15.25 | 6.00 | 5.50 |
| 23.50 | 14.75 | 6.00 | 5.16 |
| 24.00 | 14.75 | 6.00 | 5.50 |
| 24.00 | 15.00 | 6.00 | 5.50 |
| 24.00 | 15.25 | 6.00 | 5.50 |
| 24.33 | 15.25 | 5.75 | 5.50 |
| 23.83 | 15.67 | 5.83 | 5.50 |
| 24.25 | 16.25 | 6.00 | 5.50 |
| 22.50 | 14.67 | 6.00 | 5.42 |
| 24.75 | 14.50 | 6.00 | 5.67 |
| 24.00 | 15.00 | 6.00 | 5.50 |
| 23.50 | 15.16 | 6.00 | 5.58 |
| 24.00 | 15.25 | 6.00 | 5.67 |
| 23.25 | 15.33 | 6.00 | 4.83 |
| 25.00 | | 6.08 | |

Table T1. Exact growth times of each individual BN layer for DBRs C and D.

Figure 6 in the main article depicted the mapped peak reflectance level for the whole 2" DBR wafer. Here, Figure F2 depicts the wavelength for which the maximum reflectance is observed and Figure F3 presents the reflectance value for the wavelength for which the maximum is observed at the center of the wafer. As is the case with most MOCVD-grown samples, nonuniformities throughout the sample wafers are expected. This is in particular because during the process, some parts of the wafer are closer to the center of the showerhead and heater than others. This may result in a non-uniform thickness of the deposited compound throughout the substrate wafer. It can be compensated by adjusting the flow rates of reagent and carrier gases, sample-to-showerhead distance as well as heater zone factors. However, for each set of conditions and different grown materials, these values need to be adjusted separately, hence the occurrence of some kind of nonuniformities, particularly in the case of thick samples, is inevitable.

Contrary to our expectations, samples A and B fabricated using constant growth times for component BN layers exhibit less uniformity in terms of preserving the reflectance spectrum throughout the whole wafer than samples C and D, for which the growth times vary with each individual layer. Let us begin with some additional commentary on Figure 6 in the main article. In the case of DBRs A and B, the reflectance remains over 87% only in the middle inch of the wafer, then drops significantly to 80% or below at 2 mm from the edge. For samples C and D, the high peak reflectance values are measured for the whole wafer: The high-quality area extends almost to the edges of the sample as the reflectance does not drop below 85% at 2 mm from the wafer edge. This indicates that the growing boron nitride does not exhibit a constant growth rate between individual BN layers deposited at the same temperature. Rather, the deposition rate varies during the process and manual correction to the growth time of the individual layers allows to better match the $d=\lambda/4n$ condition for the optimal thickness of the DBR component layers. This will be discussed further in the text along with STEM results.

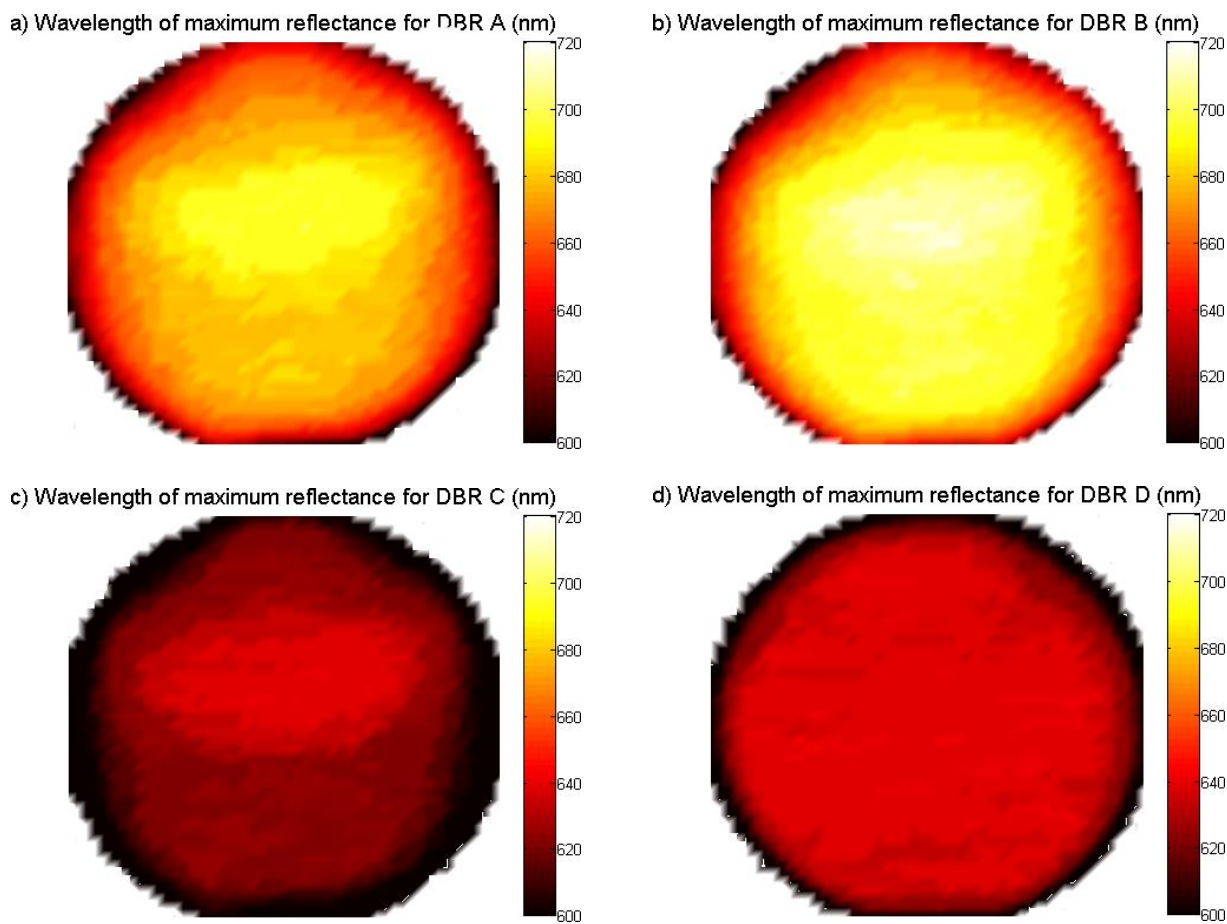

Fig. F2 The wavelength for which the peak reflectance is observed, depicted at different points of the 2" DBR samples. The upper parts of the wafers were located closer to the center of the showerhead in the MOCVD process.

Nonuniformities occur also in the case of the wavelength for which the maximum reflectance is observed. For samples A-C, the wavelength of maximum reflectance is uniform only within the area of about 1 cm × 3 cm located roughly at the center, but shifted slightly towards the parts of the wafers which were located closer to the showerhead in the MOCVD process, represented by the upper parts of graphs depicted on Fig. F2. This suggests, that even slight nonuniformities in the substrate temperature or reagent flow may influence the thickness of individual BN layers, and in turn, the structure and optical properties of the DBR. Similarly, the reflectance maps presented in Fig. F3 reflect the shapes from Fig. F2, conforming that, for a given wavelength, the spatial area of high reflectance is small and only located at the center of the wafer, shifted slightly towards the center of the showerhead.

Surprisingly, the wavelength of peak reflectance, as well as reflectance value for a single wavelength, are almost perfectly uniform for sample D, which is a DBR grown in nitrogen as a carrier gas. This might be because the sapphire substrate exhibits different surface properties when preheated at high temperatures under different conditions, as suggested in [3,4]. In our case, it is possible that ramping the sapphire substrate under nitrogen, as opposed to hydrogen, activates the surface of the substrate, allowing better adhesion of BN to the sapphire. Moreover, shorter growth times in nitrogen reduce the probability of growing three-dimensional structures, making the whole deposition process more uniform.

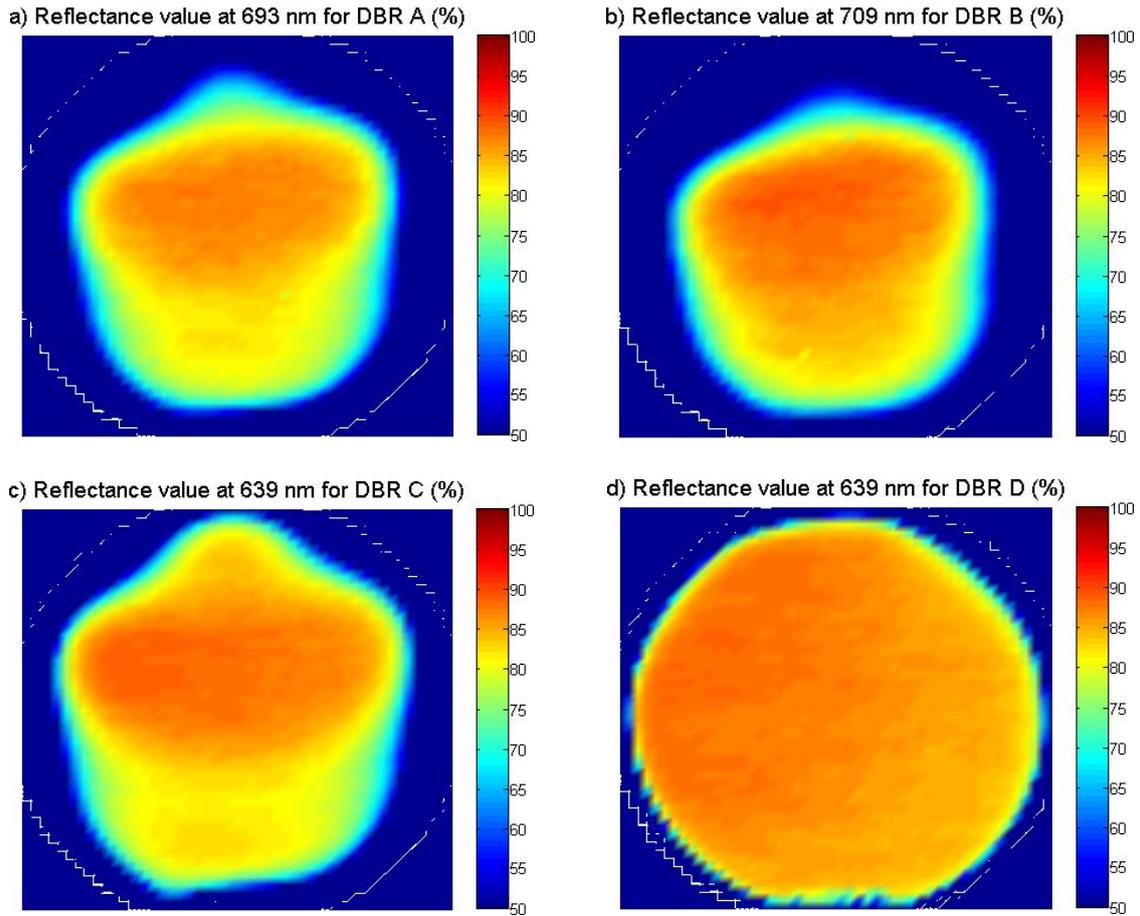

Fig. F3 Reflectance level measured at different points of the 2" DBR samples, at wavelengths for which maximum reflectance was observed at the center of the wafer. The upper parts of the wafers were located closer to the center of the showerhead in the MOCVD process.

Samples C and D exhibit remarkably similar spectral profiles of reflectance at the center of the wafer, even though they have been grown with the use of different carrier gases. This is surprising considering that BN layers grown in nitrogen are prone to incorporate large amounts of carbon. Furthermore, much faster growth rates provide greater opportunities for potential error when switching between the growth of the two types of layers.

Figure F4 presents the thicknesses of individual BN layers in DBR A determined from STEM images (Fig 7 in the main article). The values in Fig F4a are derived from Fig 7c assuming that the boundary between the high- and low-index BN layers lies in the middle of the intensity slope of the STEM signal. In that case, the thickness of the first BN layer grown at 640 °C is 89.1 nm, which is within 1.6% of the value determined by spectroscopic ellipsometry, assuming that for the single BN layer, its thickness is a linear function of growth time. For subsequent BN layers grown at 640 °C, their thickness increases even though the growth time has been kept constant. After depositing four layer pairs, the thickness of each individual BN layer grown at 640 °C stabilizes at 105.8 nm, which is 18.7% greater than the thickness of the initial layer. This indicates that for initial high-index BN layers, there is a strong interaction with the sapphire substrate, which inhibits the BN growth rate, and only after depositing 3-4 layer pairs this interaction can be neglected. In the case of BN layers grown at 820 °C, the thickness of each individual layer initially drops from 85.4 nm to 81.7 nm, but after depositing three layer pairs, it increases to 94.6 nm. The average thickness of this type of layer within the whole DBR A is 88.5 nm, which is within 1.7% of the values estimated from the optical measurements.

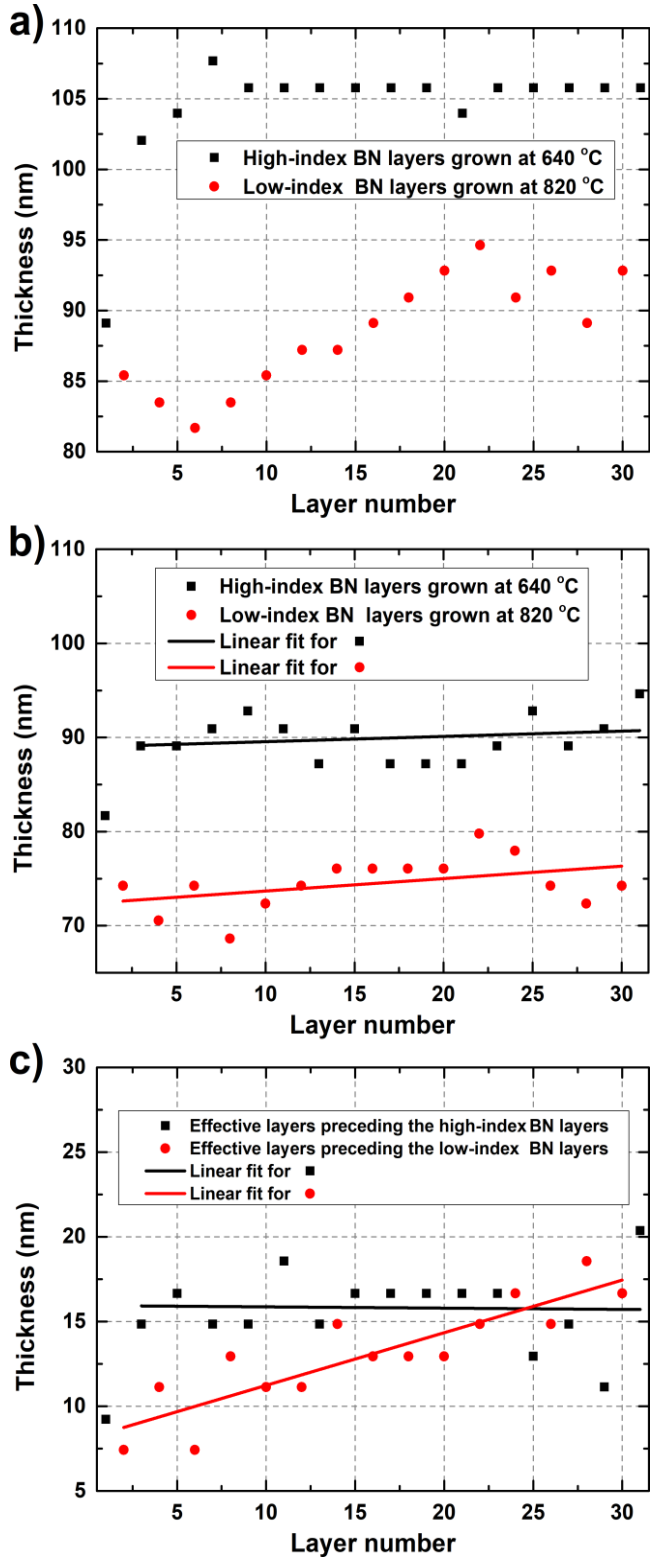

Fig F4a presents layer thickness values for a simple model where a perfect boundary between the neighboring layers is assumed. In reality, these boundaries are frequently fuzzy, and for optical simulations, we need a model which takes that into account. In such a model, we treat the high-intensity plateaus of the STEM signal in Fig. 7c in the main article as high-index BN layers, the low-intensity plateaus as low-index BN layers, and the intensity slopes as transition layers with an effective refractive index. In that case, the thicknesses of individual BN layers are presented in Fig. F4b and the thicknesses of transition layers in Fig. F4c.

Disregarding the very first high-index BN layers, for which the thickness is heavily influenced by the interaction with the substrate, the linear fits for both the high-index and low-index BN layer thicknesses exhibit an upward slope, although in both cases it is very slight – the directional coefficients are 0.056 and 0.132 for the black and red lines, respectively. The situation is quite different in regards to the effective layers. For those preceding the high-index BN layers, the fit is almost flat, with a directional coefficient of -0.008. For those preceding the low-index BN layers, the upward slope is substantial with a directional coefficient of 0.311. This means that apart from the fading of initial BN-substrate interaction, variations of the BN layer thickness can be attributed to increasing the thickness of the intermediate effective layer between the high-index and low-index BN films at the early stages of the low-index BN growth. These thickness variations are the most likely reason for the optical nonuniformities depicted in Figs F2 and F3 described in previous paragraphs.

Fig. F4 a) thicknesses of individual BN layers determined from STEM measurements, assuming that the boundary between neighboring layers lies in the middle of the intensity slope in Fig. 7c in the main article. b) thicknesses of individual BN layers assuming that only the intensity plateaus in Fig. 7c in the main article. represent the actual BN layers. c) Thicknesses of effective transition layers, assuming that they are represented by the intensity slopes in Fig. 7c in the main article. The layer number corresponds to the order in which the layers have been deposited.

Data from Fig. F4 can be verified by implementing it in a Transfer Matrix Model together with refractive index values for BN layers determined using spectroscopic ellipsometry to simulate normal reflectance as a function of wavelength. Best fits were achieved as follows:

- The model considered both the true BN layers as well as the transition layers as separate layers.
- Thicknesses for the true BN layers were taken from Fig. F4b and for transition layers from Fig. F4c
- Refractive indices of the true BN layers were taken from spectroscopic ellipsometry measurements, presented in Fig. 2 in the main article.
- Refractive indices of the transition layers were calculated using the Bruggeman formula [5] with both the high- and low-index BN layers treated as components with a 50% fill fraction.

The Bruggeman formula is as follows:

$$n_{BG}^2 = \varepsilon_{BG} = \frac{\varepsilon_1(2f_1-f_2)+\varepsilon_2(2f_2-f_1)\pm\sqrt{8\varepsilon_1\varepsilon_2(\varepsilon_1(2f_1-f_2)+\varepsilon_2(2f_2-f_1))}}{4} \quad (1)$$

Where $n_{BG}$ and $\varepsilon_{BG}$ are the complex refractive index and complex permittivity of the layer, respectively; $\varepsilon_1$ and $\varepsilon_2$ are the permittivity values of the first and second component materials; $f_1$ and $f_2$ are the volume fractions of the first and second component materials, respectively. Note that $f_1+f_2=1$. The formula does not distinguish which material acts as a medium and which as an inclusion material.

The fitting results are presented in Fig. 8 in the main article. Both for DBRs A and B the fit is very good, which yields that both the thickness values determined from the STEM measurements as well as the refractive index values determined from spectroscopic ellipsometry are correct.

The Bruggeman formula may also be used to estimate the level of porosity for the two types of fabricated BN layers. Assuming that one of the components is solid BN with permittivity values taken from the literature [6] and that the pores are filled with vacuum or air with permittivity equal to 1, we can iteratively fit the calculated refractive index curves to those extracted from spectroscopic ellipsometry. Moreover, using the Bruggeman formula along with the Percus-Yevik correction term [7,8] allows for estimating not only the level of porosity but also the pore size. The equation is then as follows:

$$\varepsilon_{BG-PY} = \varepsilon_{BG} + \frac{2if_i\varepsilon_m\pi^3 a^3}{\lambda^3} \cdot \frac{(1-f_i)^4}{(1+2f_i)^4} \cdot \left|\frac{\frac{\varepsilon_i-\varepsilon_m}{\varepsilon_i+2\varepsilon_m}}{f_i\frac{\varepsilon_i-\varepsilon_m}{\varepsilon_i+2\varepsilon_m}}\right|^2 \quad (2)$$

where, $\varepsilon_m$ is the complex permittivity of the medium (i.e. the majority material, in our case - BN), $\varepsilon_i$ is the complex permittivity of the inclusion (i.e. the minority material, in our case - pores), $f_i$ is the volume fraction of the pores and $a$ is the average radius of a pore. Results are depicted in Figure F5. Best results were achieved for 31.8% pore fraction and 55 nm pore diameter in the case of the high-index BN layer, and 43.1% pore fraction and 80 nm pore diameter in the case of the low-index BN layer.

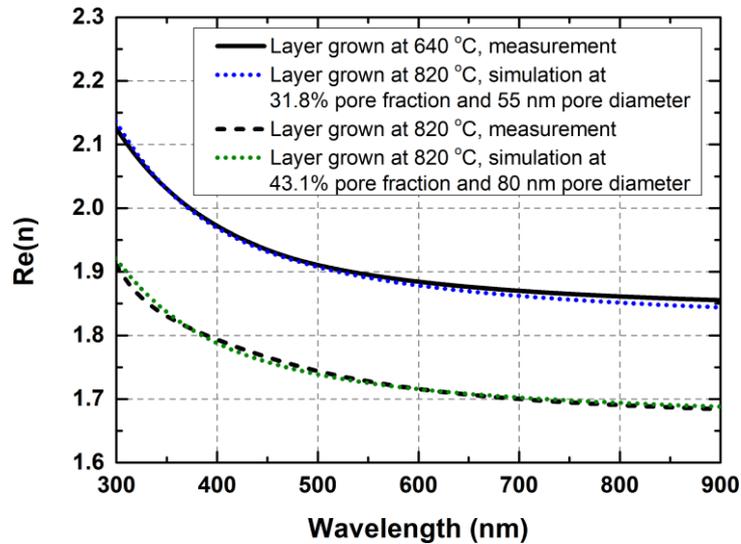

Fig. F5. Real part of the refractive index for layers grown at 640 °C and 820 °C, along with Bruggeman model simulated curves with the Percus-Yevik correction term.

Figure F6 presents reflectance maps for DBR E, similar to those presented in Fig F2 and F3. Due to the fact, that the very first BN layer was grown substantially longer than the following layers, the reflectance is much more uniform than in the case of DBRs A and B, even though the growth times for each BN layer have been fixed.

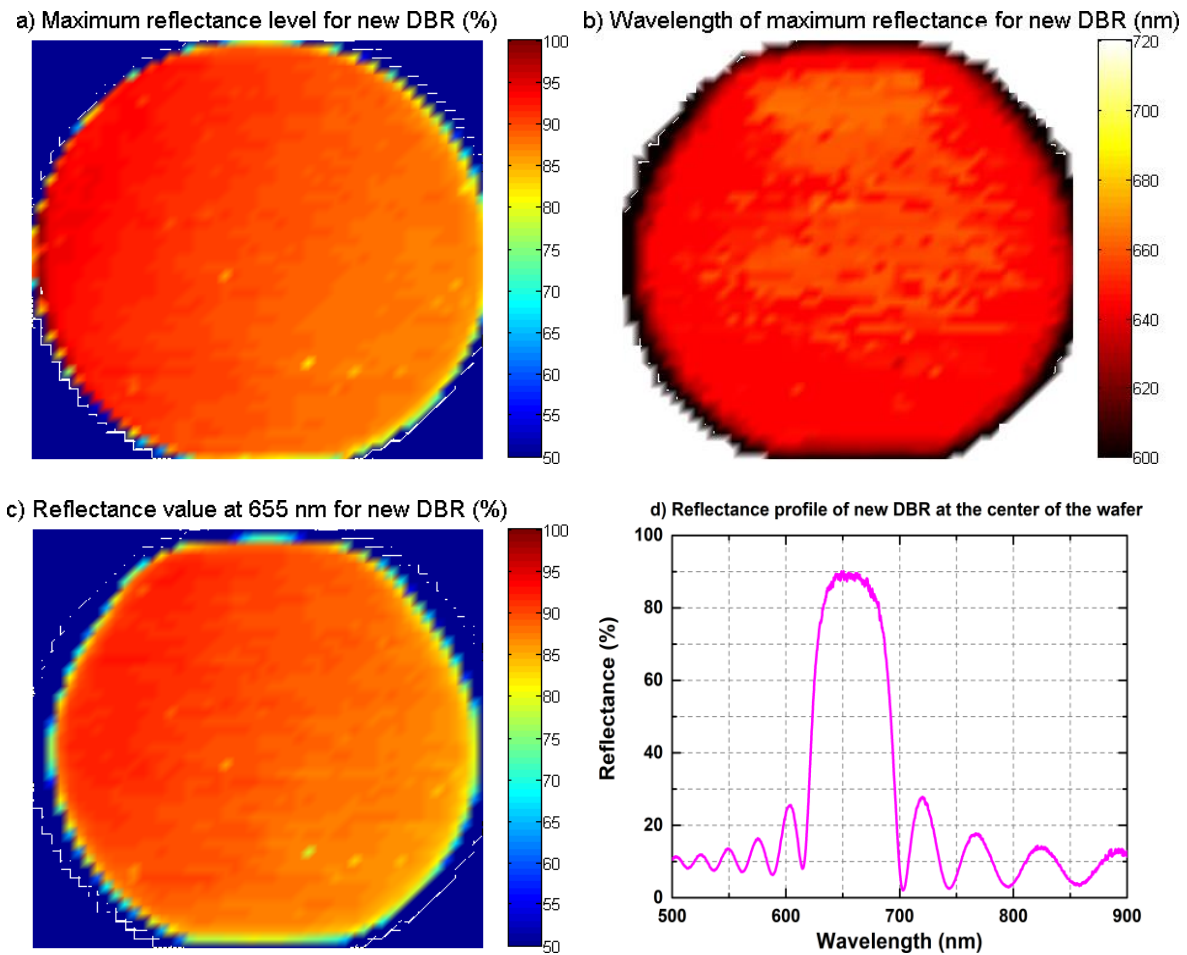

Fig. F6. a) peak reflectance level, b) the wavelength for which the peak reflectance is observed, c) reflectance level at 655 nm, measured for DBR E. The upper parts of the wafer were located closer to the center of the showerhead in the MOCVD process. d) Reflectance profile of DBR E measured at the center of the wafer.

Supplementary Video V1 is available here:
https://www.dropbox.com/s/qaypmnr43rqcpx8/Supplementary%20Material%20-%20Video%20V1.mp4?dl=0

Supplementary Video V2 is available here:
https://www.dropbox.com/s/cw3ztqvmz4cqzze/Supplementary%20Material%20-%20Video%20V2.mp4?dl=0